\documentclass[12pt]{article}
\usepackage{amsmath}
\usepackage{amssymb}
\usepackage{times}
\usepackage{epsfig,graphics}
\usepackage{color}
     \definecolor{red}{rgb}{1.0,0,0}
     \definecolor{green}{rgb}{0,1.0,0}
     \definecolor{blue}{rgb}{0,0,1.0}

\begin{document}
\pagestyle{plain}
\hsize = 6.5 in
\vsize = 8.5 in
\hoffset = -0.75 in
\voffset = -0.5 in
\baselineskip = 0.29 in
\def\vf {{\bf f}}
\def\vj {{\bf j}}
\def\vn {{\bf n}}
\def\vP{{\bf p}}
\def\vp{{\bf p}}
\def\vx{{\bf x}}
\def\vr{{\bf r}}
\def\vu{{\bf u}}
\def\vv{{\bf v}}
\def\vw {{\bf w}}
\def\vz {{\bf z}}
\def\mA{{\bf A}}
\def\mB{{\bf B}}
\def\mC{{\bf C}}
\def\mD{{\bf D}}
\def\mG{{\bf G}}
\def\mI{{\bf I}}
\def\mS{{\bf S}}
\def\mQ{{\bf Q}}
\def\mR{{\bf R}}
\def\mT{{\bf T}}
\def\mU{{\bf U}}
\def\mV{{\bf V}}
\def\mX{{\bf X}}
\def\mGa{\mbox{\boldmath$\Gamma$}}
\def\mpsi{\mbox{\boldmath$\psi$}}
\def\mPi{\mbox{\boldmath$\Pi$}}
\def\mXi{\mbox{\boldmath$\Xi$}}
\def\vg {\mbox{\boldmath$\gamma$}}
\def\vpi{\mbox{\boldmath$\pi$}}
\def\mphi{\mbox{\boldmath$\phi$}}
\def\vnu{\mbox{\boldmath$\nu$}}
\def\vkappa{\mbox{\boldmath$\kappa$}}
\def\tx{\widetilde{x}}
\def\tA{\widetilde{A}}
\def\PP{\mathbb{P}}
\def\MM{\mathbb{M}}
\def\ow{\overline{w}}
\def\tw{\widetilde{w}}
\def\wtR{R}

\def\( {\left( }
\def\) {\right) }

\title{Fitness and
Entropy Production in a Cell Population Dynamics
with Epigenetic Phenotype Switching}

\author{Hong Qian\\
\\
Department of Applied Mathematics\\
University of Washington, Seattle, WA 98195, USA
}

\maketitle

\begin{abstract}
	Motivated by recent understandings in the stochastic natures 
of gene expression, biochemical signaling, and spontaneous 
reversible epigenetic switchings, 
we study a simple deterministic cell population dynamics in which 
subpopulations grow with different rates and individual cells can 
bi-directionally switch between a small number of different 
epigenetic phenotypes. 
Two theories in the past, the population dynamics and 
thermodynamics of master equations, separatedly defined two 
important concepts in mathematical terms: the {\em fitness}
in the former and the (non-adiabatic) {\em entropy production}
in the latter.  Both play important roles in 
the evolution of the cell population 
dynamics. The switching sustains the variations among the 
subpopulation growth thus continuous natural selection. As a 
form of Price's equation, the fitness
increases with ($i$) natural selection through variations and 
$(ii)$ a positive covariance between the per capita growth and 
switching, which represents a Lamarchian-like behavior.    
A negative covariance balances  the natural selection in a 
fitness steady state | ``the red queen'' scenario.   
At the same time the growth keeps the proportions of
subpopulations away from the ``intrinsic'' switching equilibrium of 
individual cells, thus leads to a continous entropy production. A covariance, 
between the per capita growth rate and the ``chemical potential''  of
subpopulation, counter-acts the entropy production.   
Analytical results are obtained for the limiting cases of growth 
dominating switching and vice versa.
\end{abstract}

\section{Introduction}
Darwinian evolution theory states that variations among 
subpopulations contribute to the increasing {\em fitness}
of the total population.  One of the important issues in this
theory is how to maintain the variations while {\em natural
selection} continuously ``purifiying'' the fittest.  In 
population dynamics with Mendelian inheritance, this
problem has been solved by the work of Hardy and
Weinberg, and the theory of population genetics developed by
J.B.S. Haldane, R.A. Fisher, S. Wright, and many other pioneers \cite{haldane_24,fisher,wright}.   See excellent texts
\cite{crow_kimura_book,roughgarden_book,js_book} and a 
recent review \cite{ewens_qb} 
together with the references cited within.

	From a population dynamics perspective \cite{qian_nonl}, 
the key to sustain the variations among the subpopulation within a population
is to maintain a stationary frequency, or proportion, for each 
subpopulation.  In the theory of Markov processes, the concepts of
relative entropy and entropy production  are quantitative 
measures for such behavior.  In recent years, a
rather complete {\em stochastic thermodynamic theory} of 
Markov processes has emerged \cite{qqt_jsp_02,ge_qian_10,esposito,seifert_rpp,kou_qian}.   

	An integration of these two theories is a
conceptual problem for a general 
population dynamics with intra-population transformations
among subpopulations.   We note that one of the key differences
between the classical Lea-Coulson theory of bacterial
population growth with mutations \cite{zheng}
and the problem of cell population growth with epigenetic 
switching is that in the former a back mutation can be safely
neglected, while in the latter bi-directional switchings 
among a small number of possible states occur, thus the notion of
dynamical equilibrium among subpopulations becomes
meaningful in principle. Indeed  
Markov dynamics with {\em ergodic stationary 
state} among subpopulations also has emerged in laboratory studies
\cite{huang08,lander}.  Such dynamics 
occurs in the population biology of cellular epigenetic 
differentiation \cite{huang_sui,wangjin}, or certain tumor progression
\cite{huang_sui_2} and cancer cell development involving
cancer stem cells and non-stem cancer cells \cite{lander}.
According to our current understanding, biochemical regulatory
networks within an individual cell give rise to multiple attractors
which yield different phenotypic states \cite{ao_qb,qian_ge_mcb_12,wangjin_network}.  
These subpopulations of cells grow 
with different rates, but with the possibility of spontaneous
transition (switching) from one phenotypic state to another \cite{ge_qian_xie}.
In the simplest mathematical model, one has \cite{zhouda_qb}
\begin{equation}
 		\frac{dx_i}{dt} = x_ir_i(\vec{x}) +   \sum_{j=1}^n\Big(
                          T_{ij}x_j - T_{ji}x_i\Big), \ i=1,2,\cdots,n,
\label{the_eq}
\end{equation}
in which $x_i$ is the number density of the subpopulation $i$, 
$r_i$ is its per capita growth rate, and $T_{ij}$ represents the
individual cell switching rate from state $j$ to state $i$.  We
shall use the notation $\vec{x}=(x_1,x_2,\cdots,x_n)$.

	Following the pioneering work of Eigen and Schuster
\cite{eigen_schuster}, biological evolution has been 
extensively discussed in the context of genotype-to-phenotype
maps in terms of DNA sequence space and structural based
functions of proteins.  It has been recognized that there are great 
degree of degeneracy in this map \cite{vanNim}.   For microbial
population dynamics, Kussell and Leibler have studied bet-hedging 
mechanism a bacterial population with diversity can utilize when
encountering a fluctuating environment \cite{kussell_leibler}.  They 
discovered that while the information concerning the environmental 
variations can cause direct epigenetic switching in an organism,
it acts through natural selection on the organism population with 
spontaneous stochastic switchings.  More recently,
Mustonen and L\"{a}ssig developed a highly sophisticated,
time-inhomogeneous stochastic theory of fitness flux
\cite{mustonen_lassig}, which generalized Fisher's 
fundamental theorem of natural selection (FTNS). 

There is a long recognized parallelism between
the theories of population genetics and the theory of statistical
thermal physics of heterogeneous substances.  
Certainly both systems consist of a large collections of
heterogeneous individuals with random behavior, with 
emergent properties on the systems level.  This domain
of research has been very active in the 1970s, with multiple 
analogous relationships being identified and proposed.   In
fact, one of the most fascinating parallels is the 
histories of the two fields: While Boltzmann's theory
identified $\frac{dS}{dt}\ge 0$ as the Second Law 
where $S$ being the entropy of an isolated system, Eckart,
Bridgman, and Prigogine recognized that 
$dS(t) =  d_iS+d_eS$, in 
which the Second Law is represented by entropy production
$d_iS \ge 0$ while total $dS$ no longer has a definitive
sign in general.  In population genetics, Fisher's theory
identified $\frac{d}{dt}\overline{r}\ge 0$ as Darwinian
law of natural selection where $\overline{r}$ being the 
Malthusian parameter of a single locus with constant fitnesses.  
However, Price, Ewens, and Edwards later 
recognized that $d\overline{r}(t)=\partial_{ns}r +\partial_{ec}r$ \cite{price72b,edwards_94,ewens_qb},
in which FTNS is represented by  $\partial_{ns}r \ge 0$ while
total $d\overline{r}$ no longer has a definitive sign in 
general.

In the present paper, we do not attempt to deal with the
parallelism, nor important issues such as 
equilibrium and nonequilibrium statistical 
theory of evolution.  In fact, the very notions of 
``equilibrium'' and ``nonequilibrium'' defined through
traditional detailed balance can be challenged \cite{qian_pla}.
We have also neglected
many biological factors such as genetic drift
due to finite size populations.  Our goal is simply to illustrate
that Fisher's fitness in terms of the Malthusian parameter
together with relative entropy provide a quantitative 
characterization of continuous natural selection and sustained
diversity.  It is noted that, in many recent stochastic theories of
evolution, fitness of a population has been replaced
by Wright's fitness landscape, which then represented 
mathematically by a probability density function.
We only study the deterministic population
dynamics in Eq. (\ref{the_eq})
with constant per capita growth rates $r_i$ and transition rate
$T_{ij}$.  This is certainly a gross simplification; however, our aim
is to focus on the interplay between
the fitness in the population growth part and the entropy 
production in the phenotypic switching part of the dynamics.
While both are motivated from the theory of probability, 
the biological relationship between the
entropy production used in the present work and those in stochastic
theories \cite{ao_ctp,mustonen_lassig,wangjin_redqueen} 
remains to be clarified.

	Even though the issue of bi-directional switching remains 
controversial in experimental investigations and pre-clinical 
studies \cite{zhouda_qb}, there is a little double that ($i$) within
a given environment, transcription
regulations of gene expression are fundamental
biochemical processes that determine phenotypic states of a
cell; and ($ii$) biochemical reactions involved in gene regulations
are highly nonlinear with feedbacks.  Therefore, considering 
nonlinear biochemical reaction network dynamics, and its
attractors, as a basis of cellular epigenetic states is highly logical; 
and bi-directional switching
is a necessary prediction of such a mechanism.   In reality,
some transitions are fast and some transitions are slow; being
able to observe bi-directional, reversible switching or not is a matter
of time scale; and elucidating the specific molecular mechanism(s) for
a particular phenotypic switching is a matter of details.

\section{Fitness, natural selection, and entropy production}

	In this section, we first give a brief summary of the two separate
mathematical theories, one for fitness and natural selection in 
population with differential growth, and one for stochastic
thermodynamics in populations with switching transitions.

	Since the work of Haldane and Fisher more than  80 years ago \cite{haldane_24,fisher,ewens_qb}, the mean growth rate of a 
total population that consists of subpopulations with differential
growth
\begin{equation} 
            \overline{r} = \frac{\sum_{i=1}^n x_i r_i(\vec{x})}
                        {\sum_{i=1}^n x_i},
\label{eq001}
\end{equation} 
has been the mathematical quantification of  the fitness of
the population.  Note that $\overline{r}$ is actually the per capita growth rate,
i.e., the Malthusian parameter, for the entire population \cite{crow_kimura_book}:
\begin{equation}
   \overline{r}  = \frac{\displaystyle \sum_{i=1}^n x_ir_i(\vec{x})}
           {\displaystyle \sum_{i=1}^n x_i} = \frac{\displaystyle
                    \frac{d}{dt}\left(\sum_{i=1}^n x_i \right) }
          {\displaystyle \sum_{i=1}^nx_i}.
\end{equation}
We shall follow this established terminology in 
the present work.  One can also find other related mathematical 
concepts in the literature.  For examples, Wright developed the notion 
of ``fitness landscape'' \cite{wright} and Iwasa introduced a novel 
mathematical quantity called ``free fitness'' \cite{iwasa}, both were 
proposed to represent more broadly or faithfully Darwin's original 
idea and evolutionary processes \cite{johnmaddox_91}.

	For the rest of this paper, we shall assume $r_i$ to be a constant,
independent of $\{x_j\}$.  This assumption, which is not unreasonable 
for cell subpopulations growth, is essential for the mathematical
analysis carried out below.

	In the absence of switching transitions, e.g., $T_{ij}=0$, the 
change of $\overline{r}$ as a function time is
\begin{equation}
   \frac{d\overline{r}}{dt}  =
               \frac{\displaystyle \sum_{i=1}^n x_i\Big( r_i-\overline{r}\Big)^2  }
              {\displaystyle \sum_{i=1}^n x_i} = \sigma_{r}^2 \ge 0.
\end{equation}
Therefore, variations in the per capita growth rates among the 
subpopulations, $\sigma_{r}^2$, is responsible for the increase of 
fitness $\overline{r}$ by natural selection.  This simple, elegant, and 
insightful result has been known as Fisher's fundamental theorem of 
natural selection \cite{crow_kimura_book,price72b,edwards_94}.\footnote{It 
is interesting to point out that this result is intimately related to a general 
equality which plays an important role in many other branches of 
mathematics \cite{feng_kurtz}:  If $P_n(\cdot)$ is a sequence of
probability measures, for any two sets $A$ and $B$ even they are not
disjoint,
\[
\lim_{n\rightarrow\infty} \frac{1}{n}\ln P_n(A\cup B) = \max
   \left\{\lim_{n\rightarrow\infty}\frac{1}{n}\ln P_n(A), 
   \lim_{n\rightarrow\infty}\frac{1}{n}\ln P_n(B)   \right\}.
\]
}

However, in the absence of $T_{ij}$, it is also clear that 
the variation eventually disappear, and only the
fittest with the largest $r_i$ dominates.

	We now denote the frequency of the $i$th
population
\begin{equation}
    z_i = \frac{x_i}{\sum_{i=1}^n x_i}.
\end{equation}
Then Eq. (\ref{the_eq}) becomes
\begin{equation}
     \frac{dz_i}{dt} = \left(r_i-\sum_{j=1}^n r_jz_j\right)z_i
                         + \sum_{j=1}^n \Big(T_{ij}z_j-T_{ji} z_i\Big) .
\label{dzdt}
\end{equation}
Eq. (\ref{dzdt}) with all $r_i=0$ has 
a very different kind of dynamics.  Let us assume 
that $T_{ij}$ is irreducible, and $T_{ij}=0$ iff $T_{ji}=0$,
e.g., the dynamics is ``bi-directional'' with no ``absorbing state''.  Then 
there exists an unique equilibrium $\big\{z_i^*|z_i^*>0,\sum_{i=1}^n z_i^*=1\big\}$ 
such that
\begin{equation}
              \sum_{j=1}^n \Big(T_{ij}z_j^* - T_{ji}z_i^*\Big) = 0, \ 
                  \forall i.
\end{equation}
And the relative entropy
\begin{equation}
            H\big[z_i||z_i^*\big] = \sum_{i=1}^n z_i
                \ln\left(\frac{z_i}{z_i^*}\right)
\label{relent}
\end{equation}
has the important properties \cite{voigt_cmp,mackey,qian_pre_01,ao_ctp,cover_book}:
\begin{equation}
                H\big[z_i(t)||z_i^*\big] \ge 0, \ \
          \frac{d}{dt} H\big[z_i(t)||z_i^*\big] \le 0.
\end{equation}
In stochastic thermodynamics, $H$ is known as 
generalized free energy, and 
\begin{equation}
            e_p^{(na)}=-\frac{dH}{dt}
               = \sum_{i,j=1}^n T_{ji}z_i\ln
                \left(\frac{z_iz_j^*}{z_jz_i^*}\right)   \ge 0
\label{eq10}
\end{equation}
is called free energy dissipation, or non-adiabatic entropy production \cite{ge_qian_10,esposito}. In Eq. (\ref{eq10}), we used the convention
$T_{ii}=-\sum_{j=1,j\neq i}^n T_{ji}$.

The thermodynamic theory of Markov processes also identified
an adiabatic entropy production \cite{ge_qian_10,esposito} 
\begin{equation}
     e_p^{(a)} = \sum_{i,j=1}^n  T_{ij}z_j\ln 
            \left(\frac{T_{ij}z_j^*}{T_{ji}z_i^*}\right) \ge 0.
\end{equation}
We see that if $T_{ij}$ satisfy detailed balance, i.e.,
$T_{ij}z_j^*=T_{ji}z_i^*$, then $e_p^{(a)}=0$, and 
\begin{equation}
     e_p^{(na)} = \sum_{i,j=1}^n T_{ji}z_i\ln
                \left(\frac{T_{ji}z_i}{T_{ij}z_j}\right)
          = \frac{1}{2} \sum_{i,j=1}^n \Big(T_{ji}z_i-T_{ij}z_j\Big)\ln
                \left(\frac{T_{ji}z_i}{T_{ij}z_j}\right).
\label{eq0120}
\end{equation}
In L. Onsager's theory of irreversible thermodynamics of 
inanimated processes \cite{onsager}, detailed balance is a 
fundamental assumption. The right-hand-side of (\ref{eq0120}) 
is then interpreted as Onsager's ``thermodynamic flux times force''.
We do not assume detailed balance in the present work, since
for biological processes, there is clearly no fundamental reason
for such a constraint. (Such a constraint in classical physics is
due to the impossibility of perpetuate motion machine.)

	In summary, population dynamics with differential growth alone 
has increasing fitness with rate $\sigma_r^2$ that reflects Darwin's
natural selection due to
variation, and population dynamics with switching 
transformation alone has non-negative 
$e_p^{(na)}$ and $e_p^{(a)}$ reflecting the Second Law of
thermodynamics.
In Sec. \ref{sec3}, we shall study the dynamics in 
Eq. (\ref{dzdt}) that combines these two fundamental
theories.

\section{Interplay between differential growth and 
switching transitions}
\label{sec3}

If population changes following the full Eq. (\ref{dzdt}), we have a
 {\em dynamic equation for fitness}: 
\begin{eqnarray}
   \frac{d\overline{r}}{dt}  &=&  \sigma_{r}^2  -
                         \sum_{i,j=1}^n r_i
                        \Big(T_{ji}z_i -T_{ij}z_j\Big)
\\
  &=&  \sigma_{r}^2 + \overline{
             \big(r-\overline{r}\big)\big(\phi-\overline{\phi}\big)},
\label{d-law}
\end{eqnarray}
in which the last term
\begin{equation}
        \overline{
             \big(r-\overline{r}\big)\big(\phi-\overline{\phi}\big)} =
          \sum_{i=1}^n  z_i \big(r_i-\overline{r}\big)\big(\phi_i-\overline{\phi}\big),
\label{cov}
\end{equation}
with 
\begin{equation}
            \phi_i = \frac{1}{z_i}\sum_{j=1}^n  \Big(T_{ij}z_j-T_{ji}z_i\Big),
                       \textrm{ and }
             \overline{\phi} = \sum_{i=1}^n z_i\phi_i = 0.
\label{0010}
\end{equation}
$\phi_i$ is the per capita rate of increase of the $i$th subpopulation
{\em due to switching transition}.  On average $\overline{\phi}=0$.
The term in Eq. \ref{cov} can be considered as the covariance between 
the per capita growth rates due to growth and due to switching  \cite{price72a}.  
A positive covariance can be viewed phenomenologically as a 
Lamarchian selection.   Eq. (\ref{d-law})  shows that, at least in this
very simple case, one can mathematically distinguished the covariance
effect from $\sigma_r^2$, the natural selection.  Equation like (\ref{d-law})
is known as Price's theorem \cite{price72a,price72b}.

	On the other hand, from Eq. \ref{relent} we have
\begin{eqnarray}
   \frac{dH}{dt} &=&   \sum_{i,j=1}^n  \Big(T_{ji}z_i-T_{ij}z_j\Big)\ln\left(
                     \frac{z_i(t)}{z_i^*}\right) 
              + \sum_{i=1}^n \left(r_i-\sum_{j=1}^n r_jz_j\right)
								z_i\ln\left(\frac{z_i(t)}{z_i^*}\right)
\nonumber\\
	&=&  -e_p^{(na)} + \overline{\big(r-\overline{r}\big)\big(\eta-H\big) },
\label{0009}	
\end{eqnarray}
where
\begin{equation}
           \eta_i = \ln\left(\frac{z_i}{z_i^*}\right), \textrm{ and } 
          \overline{\eta} = \sum_{i=1}^n z_i\eta_i = H.
\end{equation}
The first term in (\ref{0009}) is never positive: $e_p^{(na)}\ge 0$.
The second term in (\ref{0009}) represents the covariance between the
per capita growth rate and the ``chemical potential'' of a subpopulation.
In classical statistical physics, $-\ln z^*$ and $-\ln z$ are called internal energy
and entropy, respectively.  And internal energy minus entropy is
called free energy.  It measures the deviation of $z_i$ from 
equilibrium $z_i^*$ \cite{qian_pre_01}.\footnote{Curiously, covariances also figured prominantly in 
Gibbs' theory of ensemble fluctuations.  Consider a
stochastic system with Gibbs distribution $Z^{-1}(\beta)e^{-\beta E_i}$
where $Z(\beta)=\sum_i e^{-\beta E_i}$, $E_i$ is the energy of 
of state $i$ and $\mathcal{O}_i$ is an observable associated with the 
state.  Then the ensemble average 
$\big\langle \mathcal{O} \big\rangle =$ 
$Z^{-1}(\beta) \sum_{i} \mathcal{O}_i e^{-\beta E_i}$,
whose rate of change with respect to $\beta$:
\[
         \frac{d \langle \mathcal{O}\rangle}{d\beta}
        =  - \Big\langle \big(\mathcal{O}-\langle \mathcal{O}\rangle\big)
                      \big( E- \langle E \rangle \big) \Big\rangle.
\]
}

	When the dynamics in (\ref{dzdt}) reaches stationarity, one
has $dz_i/dt=0$, i.e., $\big(r_i-\overline{r}\big)
=-\big(\phi_i-\overline{\phi}\big)$.  
In this case, $dH/dt=0$, $d\overline{r}/dt=0$,
but 
\begin{equation}
      \sigma_{r}^2 = -\overline{
             \big(r-\overline{r}\big)\big(\phi-\overline{\phi}\big)}
                      > 0,
\end{equation}
and
\begin{equation}
          e_p^{(na)} = \overline{\big(r-\overline{r}\big)\big(\eta
        -H\big) } =  -\overline{\big(\phi-\overline{\phi}\big)\big(\eta-H\big) }  > 0.
\end{equation}
The differential growth continuously keeps the population frequency
out of equilibrium, $e_p^{(na)}>0$; and an anti-Lamarchian switching
continuously keeps the variation finite, $\sigma_{r}^2>0$.
Such a nonequilibrium steady state \cite{zqq,gqq} has been termed 
``the red queen'' scenario in evolution \cite{vanvalen,ao_plr,wangjin_redqueen}.
In the simple cell dynamics, a continous evolution is necessary to oppose 
the ever-present relapsing into inferior growth states due to spontaneous 
epigenetic switching.

\section{Two asymptotic limits}

The solution to the linear Eq. \ref{the_eq} has the general form 

\begin{equation}
        \vec{x}(t) =   \Big[ e^{\mA t}\Big]\vec{x}(0),
\end{equation}
in which
\[
    \mA  = \left(\begin{array}{ccccc}
			r_1 &  0  & 0 & \dots & 0  \\
                    0   & r_2 &  0 & \dots & 0 \\
                    \vdots  & \vdots & \ddots & \vdots & 0 \\
                    0  & \dots  & \dots & r_{n-1} &  0  \\
                   0  & \dots  & \dots & 0 &  r_n 
                  \end{array} \right) + 
\]
\begin{equation} \left(\begin{array}{ccccc}
			-\sum_{j=2}^n T_{1j} &  T_{12}  & T_{13} & \dots & T_{1n}  \\[4pt]
                    T_{21}   & -\sum_{j\neq 2}T_{2j} &  T_{23} & \dots & T_{2n} \\[4pt]
                    \vdots  & \vdots & \ddots & \vdots & 0 \\[4pt]
                    T_{n-1,1}  & \dots  & \dots & -\sum_{j\neq n-1} T_{n-1,j} &  0  \\[4pt]
                   T_{n1}  & \dots  & \dots & 0 & -\sum_{j=1}^{n-1} T_{nj} 
                  \end{array} \right).
\end{equation}
We shall order the growth rates such that $r_1> r_2\ge \cdots r_n$.
In the limit of $t$ tending $\infty$, we have
\begin{equation}
     x_i(t) =   v_ie^{\lambda_{max} t}\sum_{j=1}^nu_jx_j(0),
\end{equation}
where $\lambda_{max}$ is the largest, positive eigenvalue of
$\mA$, and $\big(u_1,u_2,\cdots,u_n)$ and 
$\big(v_1,v_2,\cdots,v_n\big)^T$ are the left and right eigenvectors
corresponding to $\lambda_{max}$.
Therefore,
\begin{equation}
        z_i^{ss} =  \lim_{t\rightarrow\infty}  \frac{x_i(t)}{\sum_{i=1}^n x_i(t)} 
                     =  \frac{v_i}{\sum_{i=1}^n v_i}.
\end{equation}
When all the $r_i=0$, the $z_i^{ss}=z_i^*$.

Two special cases are particularly interesting and solvable with
the eigenvalue perturbation method.

{\bf\em Growth dominates switching.} 
When the switching rates are much smaller than 
the growth rates: $T_{ij}\ll r_k$, $i,j,k=1,2,\cdots, n$,
\begin{equation}
    \lambda_{max} \approx r_1-\sum_{j=2}^n T_{1j},
\end{equation}
and the corresponding right eigenvector $\{v_i\}$ is
\begin{equation} 
             v_i =  \frac{T_{i1}v_1}{r_1-r_i}, \ \
             v_1 =  \left(1+\sum_{i=2}^n \frac{T_{i1}}{r_1-r_i}\right)^{-1}.          
\end{equation}
Therefore,
\begin{equation}
      \overline{r} = \sum_{i=1}^n v_ir_i = \frac{\displaystyle
                    r_1+\sum_{i=2}^n\frac{r_iT_{i1}}{r_1-r_i}}
                   {\displaystyle 1+\sum_{i=2}^n \frac{T_{i1}}
                  {r_1-r_i}},
\end{equation} 
and
\begin{equation}
       \sigma_{r}^2 = \sum_{i=2}^n T_{i1}\big(r_1-r_i\big)
                  + \frac{1}{2}\sum_{i,j=2}^n \frac{T_{i1}T_{j1}\big(r_i-r_j\big)^2}
                   {(r_1-r_i)(r_1-r_j)}.
\end{equation}

	{\bf\em Switching dominates growth.}
If $r_k \ll T_{ij}$, $i,j,k=1,2,\cdots,n$. Then as the leading order
\begin{equation}
            \lambda_{max} = \sum_{i=1}^n r_iz^*_i.
\end{equation}
The corresponding right eigenvector is approximately
$z_i^*$ with a correction on the order of $\frac{r}{T}$:
\begin{equation}
             v_i = z_i^* + \xi_i, \textrm{ where } 
       \sum_{j=1}^n T_{ij}\xi_j = \big(\lambda_{max}-r_i\big)z_i^*.
\end{equation}
Since matrix $\{T_{ij}\}$ is singular with left null vector $(1,1,\cdots,1)$,
the solution $\{\xi_i\}$ is not unique.   An additional condition is
$\sum_{i=1}^n \xi_i=0$.  Hence,
\begin{equation}
     \overline{r} = \sum_{i=1}^n z_i^* r_i,
\end{equation}
\begin{equation}
    \sigma_{r}^2 = \sum_{i=1}^nz_i^*
                               \big(r_i-\overline{r}\big)^2,
\end{equation}
and
\begin{eqnarray}
     e_p^{(na)} &=& \sum_{i,j=1}^n \Big(T_{ij}v_j-T_{ji}v_i\Big)
                  \ln\left(1+\frac{\xi_i}{z_i^*}\right)
\nonumber\\
	&\approx&  \sum_{i,j=1}^n \Big(T_{ij}\xi_j
                       -T_{ji}\xi_i\Big)\left(\frac{\xi_i}{z_i^*}\right)
\nonumber\\
	&=&  \sum_{i,j=1}^n \frac{T_{ij}\xi_j\xi_i}{z_i^*} \ = \
           \sum_{i=1}^n  \xi_i\Big(\lambda_{max}-r_i\Big)
\nonumber\\
	&=& -\sum_{i=1}^n  \xi_ir_i.
\end{eqnarray}

\section{Discussion}

	Homeostatic biological processes in living organisms carry out
energy, material, and informatin transformations at the expense
of chemical free energy, e.g., negentropy \cite{schrodinger,qian_arpc}.   
Putting this intuitive 
chemical perspective aside, current theoretical understanding in entropy 
production has clearly shown its deep root in time irreversibility 
of dynamical systems \cite{zqq,gqq,seifert_rpp,qian_decomp,qian_pla}.  
According to this mathematical perspective, a non-stationary process 
with growth natually has a positive non-adiabatic entropy production
\cite{ge_qian_10,esposito,mackey}, thus a certain amount of ``free energy'' 
available for driving {\em other} dynamical processes 
out-of-equilibrium.   In the present work, we 
show that cell population growth causes a nonequilibrium
distribution among the subpopulations with respect to 
their spontaneous, epigenetic switching.   This distribution,
in return, drives a Darwinian natural selection through the 
``invisible hand'' of subpopulation variations.  Darwinian 
evolution and the Second Law of thermodynamics, thus,
are intimately coupled, at least in this simple population 
dynamics, through the certainty of mathematical analysis. 

In the past, studies on population dynamics and evolution
by natural selection were intentionally separated because 
it was felt that they are processes with widely different
time scales \cite{roughgarden_book}.  For example,
Lotka clearly introduced the notions of ``inter-species growth
dynamics'' and ``intra-species evolution'', with the former
proceeds in a rapid rate in comparison to the latter \cite{lotka_1,lotka_2}. 
Roughgarden, however, has given an illuminating discussion
on the ecological versus evolutionary times.   In particular,
he illustrated that a consideration of the latter can provide a
mechanistic understanding for the parameters in the former. 
In the language of the present work, the switching rates 
$T_{ij}$ of individual cells contribute to the overall 
growth rate of the entire population.   This is precisely the 
goal of a biochemical understanding of individual celles in a cell 
population with a diversity in epigenetic phenotypes \cite{kirschner_gerhart}.  
Being discrete nonlinear dynamic attractors \cite{kauffman}, such 
phenotypic states are naturally stable, with robustness agains both 
internal and external disturbances.  Switchings among such states,
being stochastic rare events, necessarily exhibit discontinuous jumps, e.g., 
punctuated equilibria.

\vskip 0.5cm

	I thank Drs. Ping Ao, Joe Felsenstein, Sui Huang,
Susan Rosenberg, Jin Wang, and Da Zhou for many
helpful discussions and inspirations, and my colleagues 
M. Kot, J.D. Murray, and K.K. Tung for encouragements. 
I also thank the anonymous reviewers of the paper, 
who have all given very helpful comments and suggestions.


\end{document}